\pdfoutput=1
\documentclass[preprint,10pt]{sigplanconf}
\usepackage[english]{babel}
\usepackage[T1]{fontenc}
\usepackage{amsfonts}
\usepackage{amsmath}
\usepackage{fancyvrb}
\usepackage{color}
\usepackage{units}
\usepackage[bookmarks]{hyperref}
\usepackage{mathtools}
\usepackage{amsthm}
\usepackage{bussproofs}
\usepackage{csquotes}
\usepackage{cleveref} 

\crefname{equation}{equation}{equations}

\EnableBpAbbreviations

\newcommand\real{{\ensuremath{\mathbb{R}_c}}}

\newcommand\Bool{\pmb{2}}
\newcommand\Unit{\pmb{1}}
\newcommand\Empty{\pmb{0}}
\newcommand\Sier{\Unit_\bot}
\DeclarePairedDelimiterX{\norm}[1]{\lVert}{\rVert}{#1}

\newcommand\abs[1]{| #1 |}

\newtheorem{theorem}{Theorem}[section]
\newtheorem{lemma}[theorem]{Lemma}
\newtheorem{definition}[theorem]{Definition}
\newtheorem{remark}[theorem]{Remark}
\newtheorem{corollary}[theorem]{Corollary}

\DeclareMathOperator{\Rat}{rat}
\DeclareMathOperator{\Approximation}{Approximation}
\DeclareMathOperator{\Completion}{\mathcal{C}}
\DeclareMathOperator{\Eta}{\eta}
\DeclareMathOperator{\IncreasingSequence}{IncreasingSequence}

\DeclareMathOperator{\apart}{\#}

\title{Formalising Real Numbers in Homotopy Type Theory}
\authorinfo{Ga\"etan Gilbert}{ENS Lyon}{gaetan.gilbert@ens-lyon.fr}

\begin{document}

\maketitle

\begin{abstract}
Cauchy reals can be defined as a quotient of Cauchy sequences of rationals. The limit of a Cauchy sequence of Cauchy reals is defined through lifting it to a sequence of Cauchy sequences of rationals.

This lifting requires the axiom of countable choice or excluded middle, neither of which is available in homotopy type theory. To address this, the Univalent Foundations Program uses a higher inductive-inductive type to define the Cauchy reals as the free Cauchy complete metric space generated by the rationals.

We generalize this construction to define the free Cauchy complete metric space generated by an arbitrary metric space. This forms a monad in the category of metric spaces with Lipschitz functions. When applied to the rationals it defines the Cauchy reals. Finally, we can use Altenkirch and Danielson (2016)'s partiality monad to define a semi-decision procedure comparing a real number and a rational number.

The entire construction has been formalized in the Coq proof assistant. It is available at \url{https://github.com/SkySkimmer/HoTTClasses/tree/CPP2017}.
\end{abstract}

\section{Introduction} \label{sec:introduction}

The usual process of defining the set of Cauchy real numbers proceeds in three stages: first define Cauchy sequences of rationals, then define an equivalence between Cauchy sequences, and finally quotient Cauchy sequences by the equivalence. However, proving that the so-defined Cauchy reals are Cauchy complete, i.e. that Cauchy sequences of Cauchy reals have Cauchy real limits requires the axiom of countable choice.

Alternatively, the quotient step can be replaced by working with Cauchy sequences as a setoid: this approach is used e.g. in \citet{OConnor07} which defines the completion of arbitrary metric spaces. This comes at the cost of having to make all of abstract algebra be about setoids in order to use its results for real numbers. Moreover, in the context of homotopy type theory we would like to be able to use the large amount of results about the homotopic identity but we can only do so for results about identities between bare Cauchy sequences. For instance, suppose we wish to use the principle of unique choice (which is true in homotopy type theory) to construct the unique $x : \mathbb{R}$ such that $P(x)$. Since there are multiple different Cauchy sequences representing the same real number, this will in fact only be possible if $P$ does not respect the setoid equivalence, i.e. it should be considered a property on Cauchy sequences rather than a property on real numbers.

The Higher Inductive Inductive types (HIIT) from Homotopy Type Theory \citep{HoTT} provide another construction, in only one step and without the need for an axiom of choice to prove completeness. The construction and the proof that it produces an Archimedean ordered field were outlined in the HoTT book, however formalization in the Coq proof assistant would have required workarounds for the lack of inductive-inductive types until an experimental branch by M. Sozeau started in 2015.

In \cref{sec:premetric} we define a notion of \emph{premetric space}, which on the meta level is a generalization of a metric space. From this we can define basic notions such as Lipschitz functions and limits of Cauchy sequences (or rather the equivalence but easier to work with Cauchy approximations).

\Cref{sec:cauchy-completion} generalizes the construction of the Cauchy completion of rationals from the HoTT book to arbitrary premetric spaces. This generalization shows that Cauchy completion is a monadic operator on premetric spaces (where the arrows are Lipschitz functions).

Lemmas relating to the specific structure of Cauchy reals (such as lemmas about the order on reals) are retained as shown in \cref{sec:cauchy-reals}. The monadic structure also provides a more natural way to define multiplication than that used in \citet{HoTT}.

In \cref{sec:partial-cauchy} we investigate how partial functions as per \citet{Partiality} can be defined on our definition of Cauchy reals through the example of a semi-decision procedure for the property $0 < x$.


\section{Premetric spaces} \label{sec:premetric}

We follow \citet{OConnor07} in defining distance as a relation expressing when two elements are sufficiently close. For O'Connor a metric space is a space with a relation $B_\varepsilon(x,y)$ where $x$ and $y$ are elements of the space and $\varepsilon : \mathbb{Q}^+$, which is interpreted as $d(x,y) \leq \varepsilon$.

In contrast, \citet{HoTT} defines a relation $x \approx_\varepsilon y$ for $x$ and $y$ Cauchy reals which is interpreted as $d(x,y) < \varepsilon$. We follow HoTT in using the strict order $<$.

\begin{definition}[Premetric space] \label{def:premetric}

A \emph{premetric space} is a type $A$ together with a parametric mere relation $\_ \approx_\_ \_ : \mathbb{Q}^+ \rightarrow A \rightarrow A \rightarrow Prop$ verifying the following properties:
\begin{itemize}
\item reflexivity: $\forall (\varepsilon : \mathbb{Q}^+) (x : A), x \approx_\varepsilon x$
\item symmetry: $\forall (\varepsilon : \mathbb{Q}^+) (x\; y : A), x \approx_\varepsilon y \rightarrow y \approx_\varepsilon x$
\item separatedness: $\forall x\; y : A, (\forall \varepsilon : \mathbb{Q}^+, x \approx_\varepsilon y) \rightarrow x =_A y$
\item triangularity: $\forall (x\; y\; z : A) (\varepsilon\; \delta : \mathbb{Q}^+), x \approx_\varepsilon y \rightarrow y \approx_\delta z \rightarrow x \approx_{\varepsilon + \delta} z$
\item roundedness: $\forall (\varepsilon : \mathbb{Q}^+) (x\; y : A), x \approx_\varepsilon y \leftrightarrow \exists \delta : \mathbb{Q}^+, \delta < \varepsilon \wedge x \approx_\delta y$
\end{itemize}
$\approx$ is called the closeness relation of $A$, with $x \approx_\varepsilon y$ read as "$x$ and $y$ are $\varepsilon$-close" or "the distance between $x$ and $y$ is less than $\varepsilon$".
\end{definition}

\begin{remark} \label{rmk:premetric-metric}
Classically, we can take $d(x,y) = \sup \lbrace \varepsilon : \mathbb{Q}^+, x \approx_\varepsilon y \rbrace$ with values in $\mathbb{R} + \lbrace \infty \rbrace$ to turn a premetric space into a metric space.

If we remain constructive, we expect a need for a locatedness property such as $\forall (x\; y : A)\; (q\; r : \mathbb{Q}^+), q < r \rightarrow x \approx_r y \vee x \not\approx_q y$.

We have not carried out the constructions due to lack of time, so these may not be the exact properties required. For instance without countable choice the position of the truncation may need to be different: this can be seen in \citet{HoTT} lemma 11.4.1.
\end{remark}

We now work in an arbitrary premetric space $A$.

\begin{definition}[Cauchy approximation] \label{def:approximation}
\[ \Approximation A \coloneqq \Sigma_{x : \mathbb{Q}^+ \rightarrow A} \forall \varepsilon\; \delta : \mathbb{Q}^+, x_\varepsilon \approx_{\varepsilon + \delta} x_\delta \]
A Cauchy approximation $x : \Approximation A$ can be seen as a function which given $\varepsilon$ produces a value at distance up to $\varepsilon$ of an hypothetical limit.

By abuse of notation we confuse $x : \Approximation A$ with its first projection.
\end{definition}

\begin{definition}[Limit] \label{def:islimit}
$l : A$ is a limit of the approximation $x$ when
\[ \forall \varepsilon, \delta : \mathbb{Q}^+, x_\varepsilon \approx_{\varepsilon + \delta} l \]
Since we want to express $d(x_\varepsilon,l) \leq \varepsilon$ but closeness is interpreted as $<$ we introduce an additional $\delta$.
\end{definition}

\begin{lemma} \label{lem:limit-unique}
Limits are unique: if $l_1$ and $l_2$ are limits of $x : \Approximation A$ then $l_1 = l_2$.

We may then talk about \emph{the} limit of an approximation.
\begin{proof}
By separatedness and triangularity.
\end{proof}
\end{lemma}

\begin{definition}[Cauchy completeness] \label{def:cauchycomplete}
$A$ is Cauchy complete when every Cauchy approximation has a limit. Since the limit is unique, this is equivalent to having a function
\[ lim : \Approximation A \rightarrow A \]
producing the limit for every approximation.
\end{definition}

\begin{theorem} \label{thm:q-premetric}
Rationals form a premetric space with $q \approx_\varepsilon r \coloneqq |q - r| < \varepsilon$ as its closeness.
\end{theorem}

The following lemmas make working with limits easier.

\begin{lemma} \label{lem:equiv-through-approx}
Let $y : \Approximation A$, $l_y$ and $x : A$ and $\varepsilon$ and $\delta : \mathbb{Q}^+$ such that $l_y$ is the limit of $y$ and $x \approx_\varepsilon y_\delta$. Then $x \approx_{\varepsilon + \delta} l_y$.
\begin{proof}
First strengthen the hypothesis $x \approx_\varepsilon y_\delta$ by roundedness, then finish with triangularity.
\end{proof}
\end{lemma}

\begin{lemma} \label{lem:equiv-lim-lim}
Let $x$ and $y : \Approximation A$, and $\varepsilon\; \delta\; \kappa : \mathbb{Q}^+$ such that $x_\delta \approx_\varepsilon y_\kappa$, then if $l_x$ is the limit of $x$ and $l_y$ is the limit of $y$, $l_x \approx_{\varepsilon + \delta + \kappa} l_y$.
\begin{proof}
By two applications of \cref{lem:equiv-through-approx}.
\end{proof}
\end{lemma}

\begin{lemma} \label{lem:lim-same-distance}
If $x\; y : \Approximation A$ and $\varepsilon : \mathbb{Q}^+$ are such that $\forall \delta\; \kappa : \mathbb{Q}^+, x_\kappa \approx_{\varepsilon + \delta} y_\kappa$, then for $l_x$ limit of $x$ and $l_y$ limit of $y$, $\forall \delta : \mathbb{Q}^+, l_x \approx_{\varepsilon + \delta} l_y$.

\begin{proof}
Using \cref{lem:equiv-lim-lim}, since $\varepsilon + \delta = (\varepsilon + \frac{\delta}{3}) + \frac{\delta}{3} + \frac{\delta}{3}$.
\end{proof}
\end{lemma}

\subsection{Continuity notions} \label{sec:continuity}

We will be interested in certain properties of functions between premetric spaces $A$ and $B$.

\begin{definition}[Lipschitz function] \label{def:lipschitz}
A function $f : A \rightarrow B$ is Lipschitz with constant $L : \mathbb{Q}^+$ when
\[ \forall (\varepsilon : \mathbb{Q}^+) (x, y : A), x \approx_\varepsilon y \rightarrow f\; x \approx_{L * \varepsilon} f\; y \]
If $L$ is $1$ we say that $f$ is non-expanding.
\end{definition}

\begin{definition}[Continuous function] \label{def:continuous}
A function $f : A \rightarrow B$ is continuous when
\[ \forall (\varepsilon : \mathbb{Q}^+) (x : A), \exists \delta : \mathbb{Q}^+, \forall y : A, x \approx_\delta y \rightarrow f\; x \approx_\varepsilon f\; y \]
\end{definition}

\begin{lemma} \label{lem:lipschitz-continuous}
Lipschitz functions are continuous.
\begin{proof}
Using $\delta \coloneqq \frac{\varepsilon}{L}$.
\end{proof}
\end{lemma}

Premetric spaces with continuous functions form a category.

Premetric spaces with Lipschitz functions also form a category. Notably the identity is non-expanding.

\subsection{The premetric space of functions} \label{sec:premetric-functions}

Let $A$ a type and $B$ a premetric space.

\begin{definition}[Closeness of functions] \label{def:close-arrow}
\[ f \approx_\varepsilon g \coloneqq  \exists \delta : \mathbb{Q}^+, \delta < \varepsilon \wedge \forall x : A, f\; x \approx_\delta f\; y \]
This expresses that $d(f,g) = \sup \lbrace d(f\; x, g\; x) | x : A \rbrace$.
\end{definition}

\begin{lemma} \label{lem:close-arrow-apply}
For $\varepsilon : \mathbb{Q}^+$ and $f\; g : A \rightarrow B$, if $f \approx_\varepsilon g$ then $\forall x : A, f\; x \approx_\varepsilon g\; x$.
\begin{proof}
By roundedness.
\end{proof}
\end{lemma}

\begin{theorem} \label{thm:arrow-cauchy-complete}
$A \rightarrow B$ forms a premetric space. If $B$ is Cauchy complete then so is $A \rightarrow B$, and the limit of $s : \Approximation (A \rightarrow B)$ is $\lambda y, lim\; (\lambda \varepsilon, s\; \varepsilon\; y)$.
\end{theorem}

\begin{lemma}[Limit of Lipschitz functions] \label{lem:lipschitz-lim-lipschitz}
Suppose $A$ is a premetric space and $B$ is Cauchy complete.

If $s : \Approximation (A \rightarrow B)$ is such that $\forall \varepsilon : \mathbb{Q}^+$, $s\; \varepsilon$ is Lipschitz with constant $L$, then $lim\; s$ is Lipschitz with constant $L$.
\begin{proof}
Let $\varepsilon : \mathbb{Q}^+$ and $x\; y : A$ such that $x \approx_\varepsilon y$. By roundedness there merely is $\delta\; \kappa : \mathbb{Q}^+$ such that $\varepsilon = \delta + \kappa$ and $x \approx_\delta y$.

By hypothesis $\forall \eta : \mathbb{Q}^+, s_\eta\; x \approx_{L * \delta} y$, then by roundedness $\forall \eta\; \eta' : \mathbb{Q}^+, s_\eta\; x \approx_{L * \delta + \eta'} s_\eta\; y$.

By \cref{lem:lim-same-distance} and unfolding the definition of $\lim s$ we have $\forall \eta : \mathbb{Q}^+, \lim s\; x \approx_{L * \delta + \eta} \lim s\; y$, then since $L * \varepsilon = L * \delta + L * \kappa$ we have $\lim s\; x \approx_{L * \varepsilon} \lim s\; y$.
\end{proof}
\end{lemma}

\section{Cauchy completion} \label{sec:cauchy-completion}

\subsection{Definition and eliminators} \label{sec:cauchy-complete-def}

In classical logic, we define the completion of a metric space $T$ as the quotient of the Cauchy sequences (or equivalently of Cauchy approximations) in $T$ by the equivalence $\lim f = \lim g$ (or rather an equivalent statement which doesn't assume the limit is defined). The axiom of countable choice is then used to prove that Cauchy approximations in the quotient have limits in the quotient.

Using higher inductive types, we can instead define $\Completion T$ the free complete premetric space generated by $T$. By unfolding this statement we can see what constructors it needs:
\begin{itemize}
\item generated by $T$: so there is a constructor of type $T \rightarrow \Completion T$.
\item premetric space: so we need to construct the closeness relation, and truncate $\Completion T$ to make it separated.
\item Cauchy complete: there is a constructor of type\\ $\Approximation (\Completion T) \rightarrow \Completion T$.
\end{itemize}

\begin{definition} \label{def:cauchy-completion}
$\Completion T$ has the following constructors
\begin{align*}
\Eta &: T \rightarrow \Completion T \\
\lim &: \Approximation (\Completion T) \rightarrow \Completion T
\end{align*}

The constructors of the closeness relation and the path constructors for $\Completion T$ and its closeness construct proof-irrelevant values. As such, we do not name them but instead give them as inference rules in \cref{fig:cauchy-completion}.
\begin{figure}[ht]
\begin{tabular}{c c}
\AXC{$\forall \varepsilon : \mathbb{Q}^+, x \approx_\varepsilon y$}
\UIC{$x = y$}
\DisplayProof

&

\AXC{$p, q : x \approx _\varepsilon y$}
\UIC{$p = q$}
\DisplayProof

\\\\

\AXC{$q \approx_\varepsilon r$}
\UIC{$\Eta q \approx_\varepsilon \Eta r$}
\DisplayProof

&

\AXC{$x_\delta \approx_{\varepsilon - \delta - \kappa} y_\kappa$}
\UIC{$\lim x \approx_\varepsilon \lim y$}
\DP

\\\\

\AXC{$\Eta q \approx_{\varepsilon - \delta} y_\delta$}
\UIC{$\Eta q \approx_\varepsilon \lim y$}
\DP

&

\AXC{$x_\delta \approx_{\varepsilon - \delta} \Eta r$}
\UIC{$\lim x \approx_\varepsilon \Eta r$}
\DP
\end{tabular}
\caption{Proof irrelevant constructors of $\Completion$} \label{fig:cauchy-completion}
\end{figure}

\end{definition}

We can use an explicit $fix$ expression in Coq to define the fully general induction principle with the type given in \citet{HoTT},
however it is only used through the following functions.

\begin{definition}[Simple $\Completion-$induction] \label{def:c-ind0}
Given a mere predicate $A : \Completion T \rightarrow Prop$ such that the hypotheses in \cref{fig:c-ind0} are verified,
\[ \forall x : \Completion T, A\; x \]

\begin{figure}[ht]
\begin{tabular}{c c}
\AXC{}
\UIC{$A\; (\Eta q)$}
\DP

&

\AXC{$\forall \varepsilon : \mathbb{Q}^+, A\; x_\varepsilon$}
\UIC{$A\; (\lim x)$}
\DP
\end{tabular}
\caption{Hypotheses for simple $\Completion-$induction.} \label{fig:c-ind0}
\end{figure}
\end{definition}

\begin{definition}[Simple $\approx-$induction] \label{def:equiv-rec0}
Given a mere predicate
\[ P : \mathbb{Q}^+ \rightarrow \Completion T \rightarrow \Completion T \rightarrow Prop \]
such that the hypotheses in \cref{fig:equiv-rec0} are verified,
\[ \forall \varepsilon\; x\; y, x \approx_\varepsilon y \rightarrow P\; \varepsilon\; x\; y \]

\begin{figure}[ht]
\begin{prooftree}
\AXC{$q \approx_\varepsilon r$}
\UIC{$P\; \varepsilon\; (\Eta q)\; (\Eta r)$}
\end{prooftree}

\begin{prooftree}
\AXC{$x_\delta \approx_{\varepsilon - \delta - \kappa} y_\kappa$}
\AXC{$P\; (\varepsilon - \delta - \kappa)\; x_\delta\; y_\kappa$}
\BIC{$P\; \varepsilon\; (\lim x)\; (\lim y)$}
\end{prooftree}

\begin{prooftree}
\AXC{$\Eta q \approx_{\varepsilon - \delta} y_\delta$}
\AXC{$P\; (\varepsilon - \delta)\; (\Eta q)\; y_\delta$}
\BIC{$P\; \varepsilon\; (\Eta q)\; (\lim y)$}
\end{prooftree}

\begin{prooftree}
\AXC{$x_\delta \approx_{\varepsilon - \delta} \Eta r$}
\AXC{$P\; (\varepsilon - \delta)\; x_\delta\; (\Eta r)$}
\BIC{$P\; \varepsilon\; (\lim x)\; (\Eta r)$}
\end{prooftree}
\caption{Hypotheses for simple $\approx-$induction.} \label{fig:equiv-rec0}
\end{figure}

\end{definition}

\begin{definition}[Mutual $\Completion-$recursion] \label{def:c-rec}
Let $A : Type$, a mere predicate $\sim : \mathbb{Q}^+ \rightarrow A \rightarrow A \rightarrow Prop$ and functions

\begin{gather*}
f_\eta : T \rightarrow A \\
f_{lim} : \forall (x : \Approximation (\Completion T)) (f_x : \mathbb{Q}^+ \rightarrow A),\\ (\forall (\varepsilon, \delta : \mathbb{Q}^+), f_x\; \varepsilon \sim_{\varepsilon + \delta} f_x\; \delta) \rightarrow A
\end{gather*}

If the hypotheses in \cref{fig:c-rec-hypotheses} are verified, then we have
\begin{align*}
&f : \Completion T \rightarrow A \\
&f_\approx : \forall (x, y : \Completion T) (\varepsilon : \mathbb{Q}^+), x \approx_\varepsilon y \rightarrow f\; x \sim_\varepsilon f\; y
\end{align*}
such that
\begin{align*}
&f\; (\Eta q) \coloneqq f_\eta\; q \\
&f\; (\lim x) \coloneqq f_{lim}\; x\; (f \circ x)\; \left(\lambda \varepsilon\; \delta, f_\approx\; (\varepsilon + \delta)\; x_\varepsilon\; y_\delta\right)
\end{align*}

\begin{figure}[ht]
\begin{prooftree}
\AXC{$\forall \varepsilon : \mathbb{Q}^+, x \sim_\varepsilon y$}
\UIC{$x = y$}
\end{prooftree}
\begin{tabular}{c c}
\AXC{$q \approx_\varepsilon r$}
\UIC{$f_\eta\; q \sim_\varepsilon f_\eta\; r$}
\DP

&

\AXC{$f_x\; \delta \sim_{\varepsilon - \delta - \kappa} f_y\; \kappa$}
\UIC{$f_{lim}\; x\; f_x\; H_x \sim_\varepsilon f_{lim}\; y\; f_y\; H_y$}
\DP
\end{tabular}

\begin{tabular}{c c}
\AXC{$f_\eta\; q \sim_{\varepsilon - \delta} f_y\; \delta$}
\UIC{$f_\eta\; q \sim_\varepsilon f_{lim}\; y\; f_y\; H_y$}
\DP

&

\AXC{$f_x\; \delta \sim_{\varepsilon - \delta} f_\eta\; r$}
\UIC{$f_{lim}\; x\; f_x\; H_x \sim_\varepsilon f_\eta\; r$}
\DP
\end{tabular}
\caption{Hypotheses for mutual $\Completion-$recursion.} \label{fig:c-rec-hypotheses}
\end{figure}
\end{definition}

\subsection{Properties of the completion} \label{sec:cauchy-completion-correct}

We now seek to
\begin{itemize}
\item show that $\Completion T$ is indeed a premetric space, and that $\lim$ constructs limits.
\item characterize the closeness relation: for instance $\Eta q \approx_\varepsilon \Eta r$ should be equivalent to $q \approx_\varepsilon r$.
\end{itemize}

Constructors of $\approx$ give us separatedness and proof irrelevance.

\begin{lemma}[Reflexivity] \label{lem:equiv-refl}
\[ \forall (u : \Completion T) (\varepsilon : \mathbb{Q}^+), u \approx_\varepsilon u \]
\begin{proof}
By simple induction on $u$:
\begin{itemize}
\item Let $u : T$ and $\varepsilon : \mathbb{Q}^+$. $T$ is a premetric space so $u \approx_\varepsilon u$, then $\Eta u \approx_\varepsilon \Eta u$.
\item Let $x : \Approximation (\Completion T)$ such that
	\[ \forall (\varepsilon, \delta : \mathbb{Q}^+), x_\varepsilon \approx_\delta x_\varepsilon \]
	Let $\varepsilon : \mathbb{Q}^+$. Then $x_{\varepsilon/3} \approx_{\varepsilon/3} x_{\varepsilon/3}$, so $\lim x \approx_\varepsilon \lim x$.
\end{itemize}
\end{proof}
\end{lemma}

\begin{lemma} \label{lem:c-isset}
$\Completion T$ is a set.
\begin{proof}
By \citet{HoTT} theorem 7.2.2 and separatedness.
\end{proof}
\end{lemma}

\begin{lemma}[Symmetry] \label{lem:equiv-symm}
\[ \forall (\varepsilon : \mathbb{Q}^+) (x y : \Completion T), x \approx_\varepsilon y \rightarrow y \approx_\varepsilon x \]
\begin{proof}
By simple $\approx-$induction, since $T$ has a symmetric closeness relation.
\end{proof}
\end{lemma}

To go further we need a way to deconstruct proofs of closeness. This is done by defining a function $B_\_(\_,\_) : \mathbb{Q}^+ \rightarrow \Completion T \rightarrow \Completion T \rightarrow Prop$ recursively on the two $\Completion T$ arguments which is equivalent to $\approx$.

$B$ will be defined by mutual $\Completion-$recursion as it is proof-relevant. In order to be able to prove the side conditions we will first inhabit a subtype then obtain $B$ by projection.

\begin{definition}[Concentric balls] \label{def:balls}
A set of concentric balls is a value of type
\begin{multline*}
Balls \coloneqq \Sigma_{B : \Completion T \rightarrow \mathbb{Q}^+ \rightarrow Prop} \\
	\left( \forall y\; \varepsilon, B_\varepsilon\; y \leftrightarrow \exists \delta < \varepsilon, B_\delta\; y \right) \\
	\wedge \left(\forall \varepsilon\; \delta\; y\; z, y \approx_\varepsilon z \rightarrow
		B_\delta\; y \rightarrow B_{\delta + \varepsilon}\; z
	\right)
\end{multline*}
We call the first property \emph{ball roundedness}, and the second \emph{ball triangularity}.

For $\varepsilon : \mathbb{Q}^+$ and $B_1, B_2 : Balls$, let $B_1 \approx_\varepsilon B_2$ when for $\lbrace i, j \rbrace = \lbrace 1, 2 \rbrace$
\[ \forall y\; \delta, {B_i}_\delta\; y \rightarrow {B_j}_{\delta + \varepsilon}\; y \]
\end{definition}

\begin{definition}[Upper cut] \label{def:upper-cut}
An upper cut is a predicate on $\mathbb{Q}^+$ which is upward rounded, i.e.
\[ Upper \coloneqq \Sigma_{U : \mathbb{Q}^+ \rightarrow Prop} \left( \forall \varepsilon, U_\varepsilon \leftrightarrow \exists \delta < \varepsilon, U_\delta \right) \]

For $\varepsilon : \mathbb{Q}^+$ and $U_1, U_2 : Upper$, let $U_1 \approx_\varepsilon U_2$ when for $\lbrace i, j \rbrace = \lbrace 1, 2 \rbrace$
\[ \forall \delta, {U_i}_\delta \rightarrow {U_j}_{\delta + \varepsilon} \]
\end{definition}

\begin{lemma} \label{lem:balls-separated}
The closeness on $Balls$ is separated.
\begin{proof}
Let $B^{(1)}, B^{(2)} : Balls$ such that $B^{(1)}$ and $B^{(2)}$ are $\varepsilon-$close for all $\varepsilon$. Let $\varepsilon$ and $y$, we need $B^{(1)}_\varepsilon\; y = B^{(2)}_\varepsilon\; y$. By univalence this is $B^{(1)}_\varepsilon\; y \leftrightarrow B^{(2)}_\varepsilon\; y$.

Suppose $B^{(1)}_\varepsilon\; y$, by ball roundedness there merely is $\delta < \varepsilon$ such that $B^{(1)}_\delta\; y$. $B^{(1)}$ and $B^{(2)}$ are $(\varepsilon - \delta)-$close, so we have $B^{(2)}_\varepsilon\; y$.

The second direction is the same by symmetry.
\end{proof}
\end{lemma}

\begin{lemma} \label{lem:upper-separated}
The closeness on $Upper$ is separated.
\begin{proof}
Like with \cref{lem:balls-separated} we use first roundedness then the definition of upper cut closeness at the appropriate $\varepsilon - \delta$.
\end{proof}
\end{lemma}

\begin{lemma}[Concentric balls from upper cuts] \label{lem:upper-cut-to-balls}
Suppose $B : \Completion T \rightarrow Upper$ is non-expanding, then the underlying $\Completion T \rightarrow \mathbb{Q}^+ \rightarrow Prop$ is a set of concentric balls
\begin{proof}
Ball roundedness property is exactly upper cut roundedness.

$B$ verifies ball triangularity because it is non-expanding.
\end{proof}
\end{lemma}

\begin{definition}[Balls around a base element] \label{def:equiv-alt-eta}
Let $q : T$. The set of concentric balls around $q$ is $B_\_(\Eta q,\_)$ defined by mutual $\Completion-$recursion as a non-expanding function of type $\Completion T \rightarrow Upper$ suitable for \cref{lem:upper-cut-to-balls}.

The proof relevant values are as follows:
\begin{itemize}
\item base case: $B_\varepsilon(\Eta q,\Eta r) \coloneqq q \approx_\varepsilon r$. This produces an upper cut by roundedness of $T$.
\item limit case: $B_\varepsilon(\Eta q, \lim x) \coloneqq \exists \delta < \varepsilon, B_{\varepsilon - \delta}(\Eta q, x_\delta)$. This produces an upper cut by the induction hypothesis and roundedness at the recursive call.
\end{itemize}

The remaining hypotheses expressing that the construction is non-expanding are hard to see through on paper. In Coq however reduction makes how to proceed obvious. Let us consider the $\eta-lim$ case.

Let $q\; r : T$, $\varepsilon\; \delta : \mathbb{Q}^+$ such that $\delta < \varepsilon$, and $y : \Approximation (\Completion T)$ such that we have $\lambda \kappa\; \xi, B_{\xi}(\Eta q,y_{\kappa})$. This later function is an approximation on upper cuts. Finally the induction hypothesis is that $\left( \lambda \kappa, q \approx_\kappa r \right) \approx_{\varepsilon - \delta} \left( \lambda \kappa, B_\kappa(\Eta q,x_\delta) \right)$ as upper cuts.

In that context, we need to prove that $\left( \lambda \kappa, q \approx_\kappa r \right) \approx_\varepsilon \left( \lambda \kappa, B_\kappa(\Eta q,\lim x) \right)$ as upper cuts. Let $\kappa : \mathbb{Q}^+$, we have two goals:
\begin{itemize}
\item If $q \approx_\kappa r$ then $B_{\kappa + \varepsilon}(\Eta q, \lim x)$ i.e.
	\[ \exists \delta < \kappa + \varepsilon, B_{\kappa + \varepsilon - \delta}(\Eta q, x_\delta) \]
	By the induction hypothesis and $q \approx_\kappa r$ we have
	\[ B_{\varepsilon - \delta + \kappa}(\Eta q, x_\delta) \]
	with $\delta < \varepsilon < \varepsilon + \kappa$.
\item If $\exists xi < \kappa, B_{\kappa - \xi}(\Eta q, x_\xi)$ then $q \approx_{\kappa + \varepsilon} r$.\\
	Because $\lambda \kappa\; \xi, B_{\xi}(\Eta q,y_{\kappa})$ is a cut approximation we have $B_{\kappa - \xi + \delta + \xi}(\Eta q, x_\delta) = B_{\kappa + \delta}(\Eta q, x_\delta)$. Then by induction hypothesis $q \approx_{\kappa + \varepsilon} r$.
\end{itemize}
\end{definition}

We then similarly define the concentric balls around a limit point, and show that this definition and \cref{def:equiv-alt-eta} respect $\approx$ using simple $\Completion-$induction. In order to have space for more interesting proofs we shall simply recap what results we obtain from this process.

\begin{theorem} \label{thm:equiv-alt}
We have for all $(\varepsilon : \mathbb{Q}^+)$ and $x\; y : \Completion T$, $B_\varepsilon(x, y) : Prop$ such that $\lambda x\; y\; \varepsilon, B_\varepsilon(x, y)$ is a non-expanding function from $\Completion T$ to $Balls$. Additionally we have the following computation rules:
\begin{align*}
&B_\varepsilon(\Eta q, \Eta r) \coloneqq q \approx_\varepsilon r \\
&B_\varepsilon(\Eta q, \lim y) \coloneqq \exists \delta < \varepsilon, B_{\varepsilon - \delta}(\Eta q, y_\delta) \\
&B_\varepsilon(\lim x, \Eta r) \coloneqq \exists \delta < \varepsilon, B_{\varepsilon - \delta}(x_\delta, \Eta r) \\
&B_\varepsilon(\lim x, \lim y) \coloneqq \exists \delta + \kappa < \varepsilon, B_{\varepsilon - \delta - \kappa}(x_\delta, y_\kappa)
\end{align*}
\end{theorem}

\begin{theorem} \label{thm:equiv-alt-equiv}
$B_\varepsilon(x,y)$ and $x \approx_\varepsilon y$ are equivalent.
\begin{proof}
We prove both sides of the equivalence separately:
\begin{itemize}
\item $\forall (u, v : \Completion T) (\varepsilon : \mathbb{Q}^+), B_\varepsilon(u, v) \rightarrow u \approx_\varepsilon v$\\
	By simple induction on $u$ then $v$, then using the computation rules of $B$ and the constructors of $\approx$.
\item $\forall (\varepsilon : \mathbb{Q}^+) (u, v : \Completion T), u \approx_\varepsilon v \rightarrow B_\varepsilon(u, v)$\\
	By simple $\approx-$induction, with each case being trivial.
\end{itemize}
\end{proof}
\end{theorem}

We can now use the computation rules in \cref{thm:equiv-alt} as computation rules for $\approx$.

\begin{theorem} \label{thm:c-premetric}
$\Completion T$ forms a premetric space.
\begin{proof}
Roundedness of $B$ as a closeness relation is obtained from roundedness as a function into $Balls$, then we use that $B$ equals $\approx$ to have roundedness of $\approx$.

The triangularity property of $B$ as a function into balls together with \cref{thm:equiv-alt-equiv} shows that $\approx$ is triangular.

Separatedness comes by definition of $\Completion T$, and the other properties of a premetric space are already proven in \cref{lem:equiv-symm,lem:equiv-refl}.
\end{proof}
\end{theorem}

\begin{corollary} \label{lem:eta-injective}
$\Eta$ is injective.
\begin{proof}
By separatedness.
\end{proof}
\end{corollary}

\begin{theorem} \label{thm:equiv-lim}
$\Completion T$ is Cauchy complete, i.e. for all $x : \Approximation (\Completion T)$, $\lim x$ is the limit of $x$.
\begin{proof}
\Cref{lem:equiv-through-approx} also holds for $\Completion T$:
\begin{multline*}
\forall (u : \Completion T) (y : \Approximation (\Completion T)) (\varepsilon, \delta : \mathbb{Q}^+), \\
u \approx_\varepsilon y_\delta \rightarrow u \approx_{\varepsilon + \delta} \lim y
\end{multline*}
By simple induction on $u$:
\begin{itemize}
\item Let $v : T, y : \Approximation (\Completion T)$ and $\varepsilon, \delta : \mathbb{Q}^+$ such that $\Eta v \approx_\varepsilon y_\delta$.\\
	Then by constructor $\Eta v \approx_{\varepsilon + \delta} \lim y$.
\item Let $x : \Approximation (\Completion T)$ such that (induction hypothesis)
	\begin{multline*}
	\forall (\varepsilon_0, \varepsilon, \delta : \mathbb{Q}^+) (y : \Approximation (\Completion T)), \\
	x_{\varepsilon_0} \approx_\varepsilon y_\delta \rightarrow x_{\varepsilon_0} \approx_{\varepsilon + \delta} \lim y	
	\end{multline*}
	and let $y, \varepsilon, \delta$ such that $\lim x \approx_\varepsilon y_\delta$.\\
	By roundedness, there merely exist $\kappa, \theta : \mathbb{Q}^+$ such that $\varepsilon = \kappa + \theta$ and $\lim x \approx_\kappa y_\delta$.\\
	The induction hypothesis used with $y \coloneqq x$ and reflexivity of $\approx$ gives that $\forall (\varepsilon, \delta : \mathbb{Q}^+), x_\varepsilon \approx_{\varepsilon + \delta} \lim x$ (i.e. $\lim x$ is the limit of $x$). Specifically, $x_{\theta/4} \approx_{3\theta/4} \lim x$.\\
	By triangularity, $x_{\theta/4} \approx_{3\theta/4 + \kappa} y_\delta$.\\
	By constructor $\lim x \approx_{\theta + \kappa + \delta} \lim y$.\\
	Then $\lim x \approx_{\varepsilon + \delta} \lim y$.
\end{itemize}
Then using this result and \cref{lem:equiv-symm} shows that $\lim x$ is the limit of $x$.
\end{proof}
\end{theorem}

\subsection{Monadic structure of the completion} \label{sec:c-continuity}

Continuity lets us characterize functions on $\Completion T$ based on their behaviour on the base elements $\Eta x$. If a function is sufficiently continuous, i.e. Lipschitz, we can even define its value on $\Completion T$ from its value on $T$: this turns the completion into a monad.

\begin{theorem} \label{thm:unique-continuous-extension}
Let $A$ a premetric space and $f, g : \Completion T \rightarrow A$ continuous functions such that
\[ \forall u : T, f\; (\Eta u) = g\; (\Eta u) \]
Then
\[ \forall x : \Completion T, f\; x = g\; x \]
\begin{proof}
By induction on $x$ (the desired property is a mere proposition because premetric spaces are sets). The base case is trivial.

Let $x : \Approximation (\Completion T)$ with the induction hypothesis
\[ \forall \varepsilon : \mathbb{Q}^+, f\; x_\varepsilon = g\; x_\varepsilon \]
By separatedness it suffices to prove that
\[ \forall \varepsilon : \mathbb{Q}^+, f (\lim x) \approx_\varepsilon g (\lim x) \]

Let $\varepsilon : \mathbb{Q}^+$. Continuity of $f$ and $g$ at $\lim x$ and $\varepsilon/2$ shows that there merely exist $\delta_f$ and $\delta_g : \mathbb{Q}^+$ such that
\[ \forall y : \Completion T, \lim x \approx_{\delta_f} y \rightarrow f (\lim x) \approx_{\varepsilon/2} f\; y \]
\[ \forall y : \Completion T, \lim x \approx_{\delta_g} y \rightarrow g (\lim x) \approx_{\varepsilon/2} g\; y \]

Let $\delta : \mathbb{Q}^+$ such that $\delta < \delta_f$ and $\delta_g$. By roundedness and because $\lim x$ is the limit of $x$, $\lim x \approx_{\delta_f} x_\delta$ and $\lim x \approx_{\delta_g} x_\delta$.

Then $f\; (\lim x) \approx_{\varepsilon/2} f\; x_\delta = g\; x_\delta$ and $g\; (\lim x) \approx_{\varepsilon/2} g\; x_\delta$.

By triangularity $f\; (\lim x) \approx_\varepsilon g\; (\lim x)$.
\end{proof}
\end{theorem}

Repeated application of \cref{thm:unique-continuous-extension} lets us deal with multiple variables. For instance, if $f$ and $g : \Completion T_1 \rightarrow \Completion T_2 \rightarrow A$ are continuous in both arguments (i.e. for all $x$, $f\; x$ and $g\; x$ are continuous, and for all $y$, $\lambda x, f\; x\; y$ and $\lambda x, g\; x\; y$ are continuous) and they coincide on $T_1$ and $T_2$ then they are equal.

\begin{theorem} \label{thm:lipschitz-extend}
Let $A$ a Cauchy complete premetric space and $f : T \rightarrow A$ Lipschitz with constant $L$. There exists $\overline{f} : \Completion T \rightarrow A$ Lipschitz with constant $L$ such that
\[ \forall x : T, \overline{f} (\Eta x) = f\; x \]

\begin{proof}
We define $\overline{f} : \Completion T \rightarrow A$ by mutual recursion, guaranteeing that the images of $\varepsilon$-close values are $L * \varepsilon$-close. This condition is exactly that $\overline{f}$ is Lipschitz with constant $L$.

In the base case we simply use $f$.

In the limit case, the induction hypothesis is $\overline{f_x} : \mathbb{Q}^+ \rightarrow A$ such that
\[ \forall \varepsilon, \delta : \mathbb{Q}^+, \overline{f_x}\; \varepsilon \approx_{L * (\varepsilon + \delta)} \overline{f_x}\; \delta \]
Then $\lambda \varepsilon, \overline{f_x}\; (\varepsilon/L)$ is a Cauchy approximation and we take its limit.

The coherence properties necessary for mutual recursion are easy given \cref{lem:equiv-through-approx,lem:equiv-lim-lim}.
\end{proof}
\end{theorem}

\begin{theorem} \label{thm:c-of-complete}
If $T$ is Cauchy complete then $\Completion T = T$.
\begin{proof}
The identity of $T$ is non-expanding, so it can be extended into $\overline{id_{T}} :  \Completion T \rightarrow T$.

$\overline{id_{T}} \circ \Eta_{T}$ is convertible to $id_{T}$.

$\Eta_{T} \circ \overline{id_{T}} = id_{\Completion T}$ by continuity.

Then $\overline{id_{T}}$ is an equivalence from $\Completion T$ to $T$, and by univalence they are equal.
\end{proof}
\end{theorem}

The above result uses univalence to get a strong identity result as opposed to the more common isomorphism. Note however that we do not get an isomorphism without univalence: the identity $\Eta_{T} \circ \overline{id_{T}} = id_{\Completion T}$ is proven by \cref{thm:unique-continuous-extension} at $A \coloneqq \Completion T$. To know that $\Completion T$ is a premetric space we need \cref{thm:equiv-alt}, which needs univalence to show that it preserves equality (in \cref{lem:balls-separated}).

\begin{theorem} \label{thm:c-idempotent-monad}
The Cauchy completion is an idempotent monad on the category of premetric spaces with Lipschitz functions.
\begin{proof}
Given $f : A \rightarrow B$ a Lipschitz function with constant $L$, $\Eta \circ f : A \rightarrow \Completion B$ and $\overline{\Eta \circ f} : \Completion A \rightarrow \Completion B$ are Lipschitz functions with constant $L$.

The identities about extension of identity and extension of composition are verified by continuity.

Then completion is a functor, and the previous theorem shows it is an idempotent monad.
\end{proof}
\end{theorem}

\begin{remark} \label{rmk:c-monad-uniform}
\citet{OConnor07} defines Cauchy completion as a monad on the category of metric spaces with uniformly continuous functions (with setoid identities). However the map operation requires the domain to have the additional \emph{prelength space} property (reversed triangularity: $\forall \varepsilon\; \delta\; a\; c, a \approx_{\varepsilon + \delta} c \rightarrow \exists b, a \approx_\varepsilon b \wedge b \approx_\delta c$) to be well-defined. It therefore seems that restricting the arrows to Lipschitz functions spared us from having to define and work with this property.
\end{remark}

Repeated Lipschitz extension can be applied to functions taking multiple arguments: if $f : A \rightarrow B \rightarrow T$ is Lipschitz in both arguments, the function $f_1 : A \rightarrow \Completion B \rightarrow T$ obtained by pointwise Lipschitz extension is itself a Lipschitz function into the Cauchy complete space $\Completion B \rightarrow T$.

\begin{lemma} \label{lem:lipschitz-extend-same-distance}
If $A$ is Cauchy complete and $f, g : T \rightarrow A$ are Lipschitz functions with constant $L$ and $\varepsilon : \mathbb{Q}^+$ is such that
\[ \forall (u : T) (\delta : \mathbb{Q}^+), f\; u \approx_{\varepsilon + \delta} g\; u \]
Then
\[ \forall (u : \Completion T) (\delta : \mathbb{Q}^+), \overline{f}\; u \approx_{\varepsilon + \delta} \overline{g}\; u \]
\begin{proof}
By simple induction on $u$, using \cref{lem:lim-same-distance} in the limit case.
\end{proof}
\end{lemma}

\begin{theorem}[Binary Lipschitz extension] \label{thm:lipschitz-extend-binary}
If $T$ is Cauchy complete and $f : A \rightarrow B \rightarrow T$ is such that for all $x : A$, $f\; x\; \_$ is Lipschitz with constant $L_1$ and for all $y : B$, $f\; \_\; y$ is Lipschitz with constant $L_2$, then $f$ can be extended into $\overline{\overline{f}} : \Completion A \rightarrow \Completion B \rightarrow T$ with the same Lipschitz properties and coinciding with $f$ on $\Eta$ values.
\begin{proof}
Unary Lipschitz extension gives us $f_1 \coloneqq \lambda x, \overline{f\; x} : A \rightarrow \Completion B \rightarrow T$ such that for all $x : A$, $f_1\; x\; \_$ is Lipschitz with constant $L1$.

$f_1$ is Lipschitz with constant $L_2$: let $\varepsilon : \mathbb{Q}^+$ and $x, y : A$ such that $x \approx_\varepsilon y$. We need to show that $f_1\; x \approx_{L_2 * \varepsilon} f_1\; y$, i.e. there merely exist $\delta_1, \kappa_1 : \mathbb{Q}^+$ such that
$L_2 * \varepsilon = \delta_1 + \kappa_1$ and $\forall z : B, \overline{f\; x}\; z \approx_{\delta_1} \overline{f\; y}\; z$.

By roundedness there merely exist $\delta, \kappa : \mathbb{Q}^+$ such that $\varepsilon = \delta + \kappa$ and $x \approx_\delta y$. Use $\delta_1 \coloneqq L_2 * \delta$ and $\kappa_1 \coloneqq L_2 * \kappa$.

By roundedness there merely exist $\delta', \kappa' : \mathbb{Q}^+$ such that $\delta = \delta' + \kappa'$ and $x \approx_{\delta'} y$. By \cref{lem:lipschitz-extend-same-distance} it suffices to prove
\[ \forall (z : B) (\theta : \mathbb{Q}^+), f\; x\; z \approx_{L_2 * \delta' + \theta} f\; y\; z \]
Since $f\; \_\; z$ is Lipschitz with constant $L_2$ we have
\[ f\; x\; z \approx_{L_2 * \delta'} f\; y\; z \]
then by roundedness the desired property.

$\Completion B \rightarrow T$ is Cauchy complete, so we have $\overline{\overline{f}} \coloneqq \overline{f_1} : \Completion A \rightarrow \Completion B \rightarrow T$ Lipschitz with constant $L_2$.

By \cref{lem:close-arrow-apply} we have that for all $y : \Completion B$, $\overline{\overline{f}}\; \_\; y$ is Lipschitz with constant $L_2$.

By $\Completion$-induction and \cref{lem:lipschitz-lim-lipschitz} we have that for all $x : \Completion A$, $\overline{\overline{f}}\; x\; \_$ is Lipschitz with constant $L_1$.
\end{proof}
\end{theorem}

\section{Cauchy reals} \label{sec:cauchy-reals}

We now have enough to define concepts specific to the Cauchy completion of the rationals, i.e. the Cauchy reals. Our goal is to show that they form an archimedean ordered field, a lattice, and that the closeness relation has the intended meaning $x \approx_\varepsilon y \leftrightarrow \abs{x - y} < \varepsilon$ (with absolute value of $x$ being the join of $x$ and $-x$).

Note that we use the constructive sense of ordered field, such that we have an apartness relation $x \apart y$ expressing $0 < \abs{x - y}$ and multiplicative inverse can only be applied on values apart from $0$.

\subsection{Addition and order relations} \label{sec:cauchy-reals-base}

The Cauchy reals $\real$ are the Cauchy completion of the rationals $\Completion \mathbb{Q}$. Let $\Rat : \mathbb{Q} \rightarrow \mathbb{R}_c$ be an alias for $\Eta$.

We follow \citet{HoTT} for the additive and order structure of $\real$: $0_{\real}$ is $\Rat 0_\mathbb{Q}$, $1_{\real}$ is $\Rat 0_\mathbb{Q}$, and $+$, $-$, $\cup$, $\cap$ and $\abs{\_}$ are defined by Lipschitz extension.

The HoTT book states:
\begin{displayquote}
Furthermore, the extension is unique as long as we require it to be non-expanding in each variable, and just as in the univariate case, identities on rationals extend to identities on reals. Since composition of non-expanding maps is again non-expanding, we may conclude that addition satisfies the usual properties, such as commutativity and associativity.
\end{displayquote}
This is a simple application of \cref{thm:unique-continuous-extension}. More complex uses need us to pay a little more attention to two issues:
\begin{itemize}
\item Consider transitivity of $\leq$:
	\[ \forall x\; y\; z : \real, x \cup y = y \rightarrow y \cup z = z \rightarrow x \cup z = z \]
	This cannot be directly proven by continuity as the statement of \cref{thm:unique-continuous-extension} does not allow for hypotheses which depend on the universally quantified variables.

	We can however strengthen this specific statement into one that can be solved by \cref{thm:unique-continuous-extension}: $\forall x\; y\; z : \real, x \cup ((x \cup y) \cup z) = (x \cup y) \cup z$. Doing this strengthening when we wish to use \cref{thm:unique-continuous-extension} has not been an issue, but it is unclear where it might be a problem and so should be kept in mind.
\item When showing that $\real$ is a group we need to prove $\forall x : \real, x + (- x) = 0$.

	The issue is that for a binary function $f : A \rightarrow B \rightarrow C$, knowing that for all $x$ and $y$ $\lambda y, f\; x\; y$ and $\lambda x, f\; x\; y$ are continuous is not sufficient to show that $\lambda x, f\; x\; x$ is continuous. The hypothesis we really want is that $f$ as the uncurried function from $A \times B$ to $C$ is continuous.

	If $\lambda y, f\; x\; y$ and $\lambda x, f\; x\; y$ are both Lipschitz with respective constant $L$ and $K$ then $f$ is Lipschitz with constant $L + K$, so this is not a problem when dealing with functions defined through Lipschitz extension like addition. However, showing that multiplication is continuous as an uncurried function deserves an explicit proof.
\end{itemize}

Except for those which have to do with multiplication, the proofs from \citet{HoTT} can be adapted with at most minor adjustments aside from the above remarks. Then $\real$ is a group, a lattice, $x \approx_\varepsilon y$ is equivalent to $\abs{x - y} < \varepsilon$, etc.

The book lacks the proof that $\lambda y, x + y$ preserves $<$. We show this by proving that $x < y$ if and only if there merely is $\varepsilon : \mathbb{Q}^+$ such that $x + \Rat \varepsilon \leq y$, which then allows us to use properties proven by continuity.

\begin{lemma} \label{lem:r-lt-exists-pos-plus-le}
Let $x, y : \real$ such that $x < y$. Then $\exists \varepsilon : \mathbb{Q}^+, x + \Rat \varepsilon \leq y$.
\begin{proof}
By definition of $<$ there merely are $q, r : \mathbb{Q}$ such that $x \leq \Rat q < \Rat r \leq y$. We take $\varepsilon \coloneqq r - q$.\\
$x \leq \Rat q$ so
\begin{equation*}
x + \Rat \varepsilon = \Rat (r - q) + x \leq \Rat (r - q) + \Rat q = \Rat r \leq y
\end{equation*}
\end{proof}
\end{lemma}

For the second direction, it is enough to show that $x < x + \Rat \varepsilon$. We need a helper lemma first.

\begin{lemma} \label{lem:r-le-close}
Let $\varepsilon : \mathbb{Q}^+$ and $x, y : \real$ such that $x \approx_\varepsilon y$. Then $y \leq x + \Rat \varepsilon$.
\begin{proof}
$y - x \leq \abs{x - y} < \Rat \varepsilon$ so $y \leq x + \Rat \varepsilon$.
\end{proof}
\end{lemma}

We can generalize \citet{HoTT} lemma 11.3.43:
\begin{lemma} \label{lem:r-lt-close-plus}
Let $x\; y\; z : \real$ and $\varepsilon : \mathbb{Q}^+$ such that $x < y$ and $x \approx_\varepsilon z$. Then $z < y + \Rat \varepsilon$.
\begin{proof}
There merely is $q : \mathbb{Q}$ between $x$ and $y$. By \citet{HoTT} lemma 11.3.43, $z < \Rat (q + \varepsilon) \leq y + \Rat \varepsilon$.

Note here that we cannot prove $\Rat (q + \varepsilon) < y + \Rat \varepsilon$ since $\lambda u, u + \Rat \varepsilon$ preserving $<$ is a future lemma.
\end{proof}
\end{lemma}

\begin{lemma} \label{lem:r-lt-cotrans}
$<_\real$ is cotransitive:
\[ \forall x, y, z : \real, x < y \rightarrow x < z \vee z < y \]
Note that $\vee$ is the truncated disjunction, i.e. the case distinction can only be made when proving a mere proposition.
\begin{proof}
By definition of $<$ we can reduce to the case where $x \coloneqq \Rat q$ and $y \coloneqq \Rat r$ for some $q, r : \mathbb{Q}$. Then we use simple $\Completion-$induction on $z$.

In the base case, we inherit the property from $\mathbb{Q}$.

In the limit case, we have $x : \Approximation \real$ such that (induction hypothesis)
	\[ \forall (\varepsilon : \mathbb{Q}^+) (q, r : \mathbb{Q}), q < r \rightarrow \Rat q < x_\varepsilon \vee x_\varepsilon < \Rat r \]
Let $q, r : \mathbb{Q}$ such that $q < r$. There are $q_1, r_1 : \mathbb{Q}$ such that $q < q_1 < r_1 < r$, and $\delta : \mathbb{Q}^+$ such that $\delta < q_1 - q$ and $\delta < r - r_1$.\\
Using the induction hypothesis with $\delta$ and $q_1 < r_1$ we can do a case distinction:
\begin{itemize}
\item if $\Rat q_1 < x_\delta$, we have $- x_\delta < \Rat (- q_1)$ and since $x_\delta \approx_{q_1 - q} \lim x$ and $-$ is non-expanding we have using \cref{lem:r-lt-close-plus} that $- \lim x < \Rat (- q_1 + (q_1 - q)) = \Rat (- q)$.
\item if $x_\delta < \Rat r_1$ using \cref{lem:r-lt-close-plus} we have $\lim x < \Rat (r_1 + (r - r_1)) = \Rat r$.
\end{itemize}
\end{proof}
\end{lemma}

\begin{lemma} \label{lem:r-lt-plus-pos}
For all $x : \real$ and $\varepsilon : \mathbb{Q}^+$, $x < x + \Rat \varepsilon$.
\begin{proof}
By simple $\Completion-$induction on $x$.

In the base case we inherit the result from $\mathbb{Q}$.

In the limit case, let $x : \Approximation \real$ such that (induction hypothesis)
\[ \forall \varepsilon, \delta : \mathbb{Q}^+, x_\varepsilon < x_\varepsilon + \Rat \delta \]
Let $\varepsilon : \mathbb{Q}^+$. By \cref{lem:r-lt-close-plus} and the induction hypothesis
we have $\forall \delta, \kappa : \mathbb{Q}^+, \lim x < x_\delta + \Rat (\delta + \kappa)$.\\
Using $\delta \coloneqq \varepsilon / 3$ and $\kappa \coloneqq 2 \varepsilon / 9$, by cotransitivity of $<$ (\citet{HoTT} lemma ) for $\lim x + \Rat \varepsilon$ we have either
\begin{itemize}
\item $\lim x < \lim x + \Rat \varepsilon$ as desired
\item $\lim x + \Rat \varepsilon < x_\delta + \Rat (\delta + \kappa)$, but this is absurd:\\
	By \cref{lem:r-le-close} $x_\delta \leq \lim x + \Rat (\delta + \varepsilon/9)$, then by adding $\delta + \kappa = 11\varepsilon/9$ to both sides $x_\delta + \Rat (\delta + \kappa) \leq \lim x + \Rat \varepsilon < x_\delta + \Rat (\delta + \kappa)$.
\end{itemize}
\end{proof}
\end{lemma}

We also need to prove $x \leq y$ from $\neg y < x$.

\begin{lemma} \label{lem:from-below-pr}
Real numbers can be approximated from below: let $x : \real$, then $\lambda \varepsilon : \mathbb{Q}^+, x - \Rat \varepsilon$ is an approximation with limit $x$.
\begin{proof}
\citet{HoTT} theorem 11.3.44 (expressing $x \approx_\varepsilon y$ as $\abs{x - y} < \Rat \varepsilon$) lets us reduce this to bureaucratic work.
\end{proof}
\end{lemma}

\begin{lemma} \label{lem:lipschitz-approx-lim}
Let $f : \real \rightarrow \real$ Lipschitz with constant $L$ and $x : \Approximation \real$. Then $\lambda \varepsilon, f\; x_{\varepsilon / L}$ is an approximation with limit $f\; (\lim x)$.
\begin{proof}
Easy.
\end{proof}
\end{lemma}

\begin{lemma} \label{lem:r-not-lt-le-flip}
Given $x, y : \real$, if $x < y$ is false then $y \leq x$.
\begin{proof}
Let $x, y : \real$ such that $x < y$ is false. Let $z \coloneqq x - y$. 

First note that $\forall \varepsilon : \mathbb{Q}^+, - \Rat \varepsilon < z$: let $\varepsilon : \mathbb{Q}^+$.
Since $y - \Rat \varepsilon < y$ by cotransitivity either $y - \Rat \varepsilon < x$ as desired, or $x < y$ which is absurd.

$y \leq x$ is equivalent to $0 \leq z$ i.e. $0 \cup z = z$. By \cref{lem:from-below-pr} $0 = \lim (\lambda \varepsilon, - \Rat \varepsilon)$ so by \cref{lem:lipschitz-approx-lim} $0 \cup z = \lim (\lambda \varepsilon, - \Rat \varepsilon \cup z) = \lim (\lambda \varepsilon, z) = z$.
\end{proof}
\end{lemma}

We still need to define multiplication, prove that it is continuous and behaves well regarding $<$, and show that reals apart from $0$ are invertible.

\subsection{Multiplication} \label{sec:r-ring}

Multiplication is not Lipschitz over all of $\mathbb{Q}$, so we cannot simply use Lipschitz extension. The definition in \citet{HoTT} first defines squaring and uses the identity $u * v = \frac{(u+v)^2 - u^2 - v^2}{2}$ to define multiplication from it. We stay closer to simple Lipschitz extension by defining multiplication on bounded intervals then joining these to cover $\real$.

\begin{definition}[Definition by surjection] \label{def:def-by-surjection}
Let $A\; B$ and $C$ sets, and $f : A \rightarrow C$ and $g : A \rightarrow B$ functions such that $g$ is a surjection and $f$ respects $\sim_g$ the equivalence relation on $A$ induced by $g$.

Then $B$ is equivalent to $A/\sim_g$ the quotient of $A$ by $\sim_g$ and there is a function $f_{\sim_g} : A/\sim_g \rightarrow C$ acting like $f$.

Composing $f_{\sim_g}$ with the equivalence defines the function $\overline{f}_{\sim_g} : B \rightarrow C$ such that $\forall x : A, \overline{f}_{\sim_g}\; (g\; x) = f\; x$.
\end{definition}

\begin{definition}[Intervals] \label{def:interval}
For $a, b : \mathbb{Q}$ (resp. $a, b : \real$), the interval space $[a, b] \coloneqq \Sigma_{x} a \leq x \leq b$ inheriting the closeness relation from the first projection forms a premetric space.

For $x : \mathbb{Q}$ (resp. $x : \real$), $a \leq a \cup (x \cap b) \leq b$ so we can define $[x]_{a,b} : [a, b]$. If $a \leq x \leq b$ then $[x]_{a,b}$ has its first projection equal to $x$.
\end{definition}

\begin{definition}[Left multiplication by a rational] \label{def:qrmult}
For any $q : \mathbb{Q}$, $\lambda r : \mathbb{Q}, q * r$ is Lipschitz with constant $q + 1$, so we define $\lambda (q : \mathbb{Q}) (y : \real), q * y$ by Lipschitz extension with constant $a$.
\end{definition}

\begin{definition}[Bounded multiplication] \label{def:r-bounded-mult}
For $a : \mathbb{Q}^+$ and $y : [- \Rat a, \Rat a]$ we define $\lambda x : \real, x *_a y$ by Lipschitz extension.
\begin{proof}
We need to check that $\lambda q : \mathbb{Q}, q * y$ is Lipschitz with constant $a$. Using \citet{HoTT} theorem 11.3.44 it suffices to show that for $x : \real$ such that $\abs{x} \leq \Rat a$ we have $\forall q\; r : \mathbb{Q}, \abs{q * x - r * x} \leq \Rat (\abs{q - r} * a)$. This is obtained by continuity.
\end{proof}
\end{definition}

\begin{lemma} \label{lem:r-qpos-bounded}
Cauchy reals are bounded by rationals, i.e. for all $x : \real$ there merely is $q : \mathbb{Q}^+$ such that $\abs{x} < \Rat q$.
\begin{proof}
By simple $\Completion-$induction on $x$.

In the base case we take $q \coloneqq \abs{x} + 1$.

In the limit case, where $x$ is $\lim f$, by the induction hypothesis there merely is $q : \mathbb{Q}^+$ such that $\abs{f\; 1} < \Rat q$.\\
$\abs{f\; 1} \approx_2 \abs{x}$ so $x < \Rat (q + 2)$.
\end{proof}
\end{lemma}

\begin{lemma} \label{lem:interval-back}
Let the following function be defined by the obvious projections:
	\[ \lbrace \_ \rbrace : \Sigma_{a : \mathbb{Q}^+} [- \Rat a, \Rat a] \rightarrow \real \]
It is surjective and respects bounded multiplication, i.e.
\[ \forall x, y, z, \lbrace x \rbrace = \lbrace y \rbrace \rightarrow z *_{x_1} x_2 = z *_{y_1} y_2 \]
\begin{proof}
The function is surjective because reals are bounded by rationals. It respects bounded multiplication by continuity.
\end{proof}
\end{lemma}

\begin{definition}[Multiplication] \label{def:r-mult}
For $x : \real$ we define $\lambda y : \real, x * y$ from
	\[ \lambda y : \Sigma_{a : \mathbb{Q}^+} [- \Rat a, \Rat a], x *_{y_1} y_2 \]
and surjectivity of $\lbrace \_ \rbrace$.
\end{definition}

Multiplication is now defined, with the following properties by definition:

\begin{lemma} \label{lem:r-mult-interval-proj-applied}
For $x : \real$ and $a : \mathbb{Q}^+$ and $y : [- \Rat a, \Rat a]$
\[ x * y_1 = x *_a y \]
\begin{proof}
By unfolding \cref{def:def-by-surjection}.
\end{proof}
\end{lemma}

\begin{lemma} \label{lem:r-mult-rat-rat}
Multiplication computes on rationals:
  \[ \forall q, r : \mathbb{Q}, \Rat q * \Rat r \equiv \Rat (q * r) \]
\begin{proof}
Checking a conversion is decidable so this proof is left as an exercise to the reader.
\end{proof}
\end{lemma}

We now need to show that multiplication is continuous as an uncurried function.

\begin{lemma} \label{lem:r-mult-lipschitz-aux-alt}
For $a : \mathbb{Q}^+$ and $y : \real$ such that $\abs{y} \leq \Rat a$, $\lambda x : \real, x * y$ is Lipschitz with constant $a$.
\begin{proof}
Using \cref{lem:r-mult-interval-proj-applied,def:r-bounded-mult}.
\end{proof}
\end{lemma}

\begin{lemma} \label{lem:r-mult-continuous-r}
For all $y : \real$, $\lambda x : \real, x * y$ is continuous.
\begin{proof}
Let $y : \real$, there merely is $a : \mathbb{Q}^+$ such that $\abs{y} \leq \Rat a$. By \cref{lem:r-mult-lipschitz-aux-alt} $\lambda x : \real, x * y$ is Lipschitz with constant $a$ and therefore continuous.
\end{proof}
\end{lemma}

\begin{lemma} \label{lem:r-mult-rat-l}
For $q : \mathbb{Q}$ and $x : \real$, $\Rat q * x = q * x$.
\begin{proof}
Using \cref{lem:r-mult-interval-proj-applied} for some $a$ bounding $x$.
\end{proof}
\end{lemma}

\begin{lemma} \label{lem:r-mult-abs-l}
Multiplication and negation distribute inside the absolute value: 
\[ \forall a, b, c : \real, \abs{a * b - a * c} = \abs{a} * \abs{b - c} \]
\begin{proof}
We can reduce to the case where $a$ is rational by continuity, then use \cref{lem:r-mult-rat-l} to replace real to real multiplication with rational to real multiplication and finish by continuity.
\end{proof}
\end{lemma}

\begin{lemma} \label{lem:r-mult-le-compat-abs}
Multiplication is compatible with $\leq$ under absolute value: for $a, b, c, d : \real$, if $\abs{a} \leq \abs{c}$ and $\abs{b} \leq \abs{d}$ then $\abs{a} * \abs{b} \leq \abs{c} * \abs{d}$.
\begin{proof}
Again we use continuity to reduce $a$ and $c$ (the variables appearing to the left of the multiplications) to their rational case, then rewrite the desired property to use multiplication of a rational and a real and finish with continuity.
\end{proof}
\end{lemma}

\begin{theorem} \label{thm:r-mult-continuous}
Multiplication is continuous as a function of 2 variables, i.e. given $u_1$ and $v_1 : \real$ and $\varepsilon : \mathbb{Q}^+$ there merely exists $\delta : \mathbb{Q}^+$ such that for all $u_2$ and $v_2 : \real$, if $u_1 \approx_\delta u_2$ and $v_1 \approx_\delta v_2$ then $u_1 * v_1 \approx_\varepsilon u_2 * v_2$.
\begin{proof}
Let $u_1, v_1 : \real$ and $\varepsilon : \mathbb{Q}^+$. There merely is $\delta : \mathbb{Q}^+$ such that $\abs{u_1} < \Rat \delta$ and $\abs{v_1} < \Rat \delta$. Let $\kappa \coloneqq \delta + 1$, then in the lemma's statement we take $\delta \coloneqq 1 \cap \frac{\varepsilon}{2(\kappa + 1)}$.

Let $u_2, v_2 : \real$ such that
\begin{itemize}
\item $u_1 \approx_{1 \cap \frac{\varepsilon}{2(\kappa + 1)}} u_2$
\item $v_1 \approx_{1 \cap \frac{\varepsilon}{2(\kappa + 1)}} v_2$
\end{itemize}
Then:
\begin{itemize}
\item $u_1 * v_1 \approx_{\varepsilon / 2} u_2 * v_1$:
	
	$\abs{v_1} \leq \Rat \delta$ so $\lambda y : \real, y * v_1$ is Lipschitz with constant $\delta$ and it suffices to prove $u_1 \approx_{\varepsilon/2\delta} u_2$.
	
	This is true from roundedness and the first $\approx$ hypothesis since $ 1 \cap \frac{\varepsilon}{2(\kappa + 1)} \leq \varepsilon/2\delta $.
\item $u_2 * v_1 \approx_{\varepsilon / 2} u_2 * v_2$:
	
	By \citet{HoTT} theorem 11.3.44 we look to prove $\abs{u_2 * v_1 - u_2 * v_2} = \abs{u_2} * \abs{v_1 - v_2} < \varepsilon/2$.
	
	In fact we have
	\begin{itemize}
	\item $\abs{u_2} \leq \abs{\kappa} = \kappa$ since $\abs{u_1} \leq \kappa$ and $u_1 \approx_1 u_2$.
	\item $\abs{v_1 - v_2} \leq \abs{\frac{\varepsilon}{2(\kappa+1)}} = \frac{\varepsilon}{2(\kappa+1)}$ since $\abs{v_1 - v_2} < 1 \cap \frac{\varepsilon}{2(\kappa + 1)}$.
	\end{itemize}
	Then by \cref{lem:r-mult-le-compat-abs} we have $\abs{u_2} * \abs{v_1 - v_2} \leq \abs{\kappa} * \abs{\frac{\varepsilon}{2(\kappa+1)}} = \varepsilon/2 * \frac{\kappa}{\kappa+1} < \varepsilon/2$.
\end{itemize}
By triangularity $u_1 * v_1 \approx_\varepsilon u_2 * v_2$.
\end{proof}
\end{theorem}

This is enough to show that $\real$ forms a partially ordered ring, but we still need to link multiplication and $<$.

\begin{lemma} \label{lem:r-mult-pos}
Multiplication of positive values produces a positive value: let $x, y : \real$ such that $0 < x$ and $0 < y$, then $0 < x * y$.
\begin{proof}
Let $x, y : \real$ such that $0 < x$ and $0 < y$, then there merely are $\varepsilon, \delta : \mathbb{Q}^+$ such that $\varepsilon < x$ and $\delta < y$.

By continuity multiplication preserves $\leq$ for nonnegative values, so $0 < \Rat (\varepsilon * \delta) \leq x * y$.
\end{proof}
\end{lemma}

\begin{lemma} \label{lem:r-mult-pos-decompose-nonneg}
For $x, y : \real$, if $0 \leq x$ and $0 < x * y$ then $0 < y$.
\begin{proof}
There merely is $\varepsilon : \mathbb{Q}^+$ such that $\Rat \varepsilon < x * y$.
By \cref{lem:r-qpos-bounded} there merely is $\delta : \mathbb{Q}^+$ such that $\abs{x} < \Rat \delta$. Then it suffices to prove $0 < \varepsilon / \delta \leq y$.

We do this using \cref{lem:r-not-lt-le-flip}: suppose $y < \varepsilon / \delta$. Since $0 \leq y$ (if $y < 0$ then $x * y \leq 0$ which is absurd), $x * y \leq \abs{x} * y \leq \varepsilon < x * y$ which is absurd.
\end{proof}
\end{lemma}

\subsection{Multiplicative inverse}

The multiplicative inverse for $\mathbb{Q}$ is Lipschitz on intervals $[\varepsilon, +\infty]$ for $\varepsilon : \mathbb{Q}^+$. We use this to extend it to positive reals, then to reals apart from $0$ using negation.

\begin{definition} \label{def:bounded-inverse}
For $\varepsilon : \mathbb{Q}^+$ the function $\lambda q : \mathbb{Q}, \frac{1}{\varepsilon \cup q}$ is defined and Lipschitz with constant $\varepsilon^{-2}$.

Then for $x : \Sigma_{\varepsilon : \mathbb{Q}^+, x : \real} \Rat \varepsilon \leq x$ we define
\[ /_\Sigma x \coloneqq \left(\overline{\lambda q : \mathbb{Q}, \frac{1}{x_\varepsilon \cup q}}\right) x_x \]
\end{definition}

\begin{definition} \label{def:r-recip}
We define the inverse of positive reals by surjection (\cref{def:def-by-surjection}) using $/_\Sigma$ and the obvious surjection from $x : \Sigma_{\varepsilon : \mathbb{Q}^+, x : \real} \Rat \varepsilon \leq x$ to $\Sigma_{x : \real} 0 < x$.

For negative values we use the identity $\frac{1}{x} \coloneqq - \frac{1}{- x}$.

This gives $\frac{1}{x}$ for any $x$ such that $x \apart 0$.
\end{definition}

\begin{lemma} \label{lem:r-recip-rat}
$\forall q : \mathbb{Q}, \Rat q \apart 0 \rightarrow \frac{1}{\Rat q} = \Rat (\frac{1}{q})$
\begin{proof}
The negative case is easily reduced to the positive case.

In the positive case there merely are $r, s : \mathbb{Q}$ such that $0 \leq r < s \leq q$, then $\frac{1}{\Rat q}$ reduces to $\Rat \frac{1}{q \cup s}$ which is equal to $\Rat \frac{1}{q}$ since $s \leq q$.
\end{proof}
\end{lemma}

\begin{lemma} \label{lem:r-recip-upper-recip}
For $x : \real$ and $\varepsilon : \mathbb{Q}^+$ such that $\Rat \varepsilon \leq x$, $\frac{1}{x} = \left(\overline{\lambda q : \mathbb{Q}, \frac{1}{\varepsilon \cup q}}\right) x$.
\begin{proof}
Easy.
\end{proof}
\end{lemma}

\begin{lemma} \label{lem:r-recip-inverse}
$\forall x : \real$, if $x \apart 0$ then $x * \frac{1}{x} = 1$.
\begin{proof}
We can reduce to the case where $0 < x$. Then there merely is $\varepsilon : \mathbb{Q}^+$ such that $\Rat \varepsilon \leq x$.

By continuity $x * \left(\overline{\lambda q : \mathbb{Q}, \frac{1}{\varepsilon \cup q}}\right) x = 1$ for all $x$ such that $\Rat \varepsilon \leq x$, and by definition, $\frac{1}{x} = \left(\overline{\lambda q : \mathbb{Q}, \frac{1}{\varepsilon \cup q}}\right) x$. 
\end{proof}
\end{lemma}

Together with the results from \citet{HoTT} section 11.3.3 we now have all results needed for $\real$ to form an Archimedean ordered field as desired.

\section{A partial function on Cauchy reals} \label{sec:partial-cauchy}

Without additional axioms, we can't define any non-constant function from $\real$ to booleans $\mathbb{B}$. In other words, no non-trivial property on $\real$ is decidable.

However we can encode non-termination as an effect in the \emph{partiality monad}, where the type of computations producing values of type $A$ is denoted $A_\bot$. Then we can define a function $isPositive : \real \rightarrow 2_\bot$ which produces $true$ on positive reals, $false$ on negative reals and does not terminate on $0$.

\subsection{The partiality monad} \label{sec:def-partiality}

In \citet{Partiality}, Altenkirch and Danielsson define the type $A_\bot$ of computations producing values of type $A$ as a HIIT. They implemented it in Agda and proved certain properties such as the existence of fixpoints and that it forms the free $\omega-$CPO on $A$.

\begin{definition}[Increasing sequences] \label{def:increasing-sequence}
\[ \IncreasingSequence A \coloneqq \Sigma_{f : \mathbb{N} \rightarrow A} \forall n, f_n \leq f_{S n} \]
As with Cauchy approximations we confuse\\ $f : \IncreasingSequence A$ with the underlying function in our notations.
\end{definition}

\begin{definition} \label{def:partial}
Given $A$ a type, the type $A_\bot$ is defined simultaneously with its order. It has the following constructors:
\begin{itemize}
\item $\Eta : A \rightarrow A_\bot$
\item $\bot : A_\bot$
\item $sup : \IncreasingSequence A_\bot \rightarrow A_\bot$
\end{itemize}
with a path constructor of type $\forall x, y : A_\bot, x \leq y \rightarrow y \leq x \rightarrow x = y$.

The order has constructors of types
\begin{itemize}
\item $\forall x : A_\bot, x \leq x$
\item $\forall x : A_\bot, \bot \leq x$
\item $\forall f, x, sup\; f \leq x \rightarrow \forall n, f_n \leq x$
\item $\forall f, x, (\forall n, f_n \leq x) \rightarrow sup\; f \leq x$
\end{itemize}
and is truncated to be propositional.

As with the Cauchy completion we have simple induction on values and simple induction on the auxiliary relation $\leq$ to prove inhabitedness of propositional types depending on computations, and non-dependent mutual recursion to define values from computations.
\end{definition}

Altenkirch and Danielsson suggest a way of defining $isPositive : \mathbb{R}^q \rightarrow \Bool_\bot$, where $\mathbb{R}^q$ is the quotient of Cauchy sequences of $\mathbb{Q}$ by the appropriate equivalence.

They first define it on Cauchy sequences of $\mathbb{Q}$ using the fixpoint operator provided by the partiality functor, then show that it respects the equivalence and extend it to the quotient $\mathbb{R}^q$.

We could not adapt that definition for the HIIT Cauchy real numbers. However, an alternate definition is possible:
\begin{itemize}
\item For $P : Prop$, $\left( \Sigma_{p : \Sier} p = \Eta \star \leftrightarrow P \right)$ is propositional.

	We can use simple $\Completion-$induction to define $p$ for all $P \coloneqq x < \Rat q$.
\item From $p$ and $q : \Sier$ such that $p$ and $q$ are not both $\Eta \star$, we define $interleave\; p\; q : \Bool_\bot$ indicating which if any is $\Eta \star$.
\item We interleave the values defined from $- x < 0$ and $x < 0$ to define $isPositive\; x$.
\end{itemize}

We assume the properties of $A_\bot$ for arbitrary $A$ from \citet{Partiality}. Let us then focus on the properties of $\Sier$.

\subsection{The Sierpinski space}

If $A_\bot$ is the type of possibly non-terminating computations returning a value of type $A$, then $\Sier$ is the type of semi-decision procedures: $p : \Sier$ semi-decides all propositions equivalent to $p = \Eta \star$.

\begin{definition} \label{def:sier-top}
$\Sier$ has a greatest element $\top \coloneqq \Eta \star$.
\begin{proof}
$\forall x : \Sier, x \leq \top$ by simple induction on $x$.
\end{proof}
\end{definition}

We can interpret $p : \Sier$ as the proposition $p = \Eta \star$ (equivalently, $\Eta \star \leq p$).

Then trivially $\top \leftrightarrow \Unit$, $\bot \leftrightarrow \Empty$.

\begin{lemma} \label{lem:sier-le-imply}
For all $a\; b : \Sier$, $a \leq b$ if and only if $a \rightarrow b$.
\begin{proof}
\item if $a \leq b$ then $a \rightarrow b$: suppose $a$, i.e. $\top \leq a$. Then $\top \leq a \leq b$, i.e. $b$.
\item if $a \rightarrow b$ then $a \leq b$: by simple induction on $a$, each case being trivial.
\end{proof}
\end{lemma}

We can also interpret $\vee$ into $\Sier$ (and $\wedge$, but we do not need it for $isPositive$).

\begin{definition}[Join on $\Sier$] \label{def:sier-join}
\begin{proof}
We define an auxiliary function by mutual recursion: for all $y : \Sier$ there is $\cup_y : \Sier \rightarrow \Sigma_{z : \Sier} y \leq z$, then $x \cup y \coloneqq (\cup_y\; x)_1$. Then $x \cup y$ is the first projection of $\cup_y\; x$. It computes as follows:

\begin{itemize}
\item $\top \cup\; y \coloneqq \top$
\item $\bot \cup\; y \coloneqq y$
\item $(\sup f) \cup\; y \coloneqq \sup (\lambda n, f_n \cup\; y)$
\end{itemize}
The proofs of the required properties are trivial. Note that we need the auxiliary function as we need a proof that $\forall x : \Sigma_{z : \Sier} y \leq z, \cup_y\; \bot = y \leq x_1$.
\end{proof}
\end{definition}

\begin{lemma} \label{lem:sier-join-semilattice}
$x \cup y$ is the least upper bound of $x$ and $y$. Then $\cup$ is a monoid operator with identity element $\bot$.
\begin{proof}
By definition and simple inductions.
\end{proof}
\end{lemma}

\begin{lemma} \label{lem:sier-join-disj}
For all $a\; b : \Sier$, $a \cup b$ if and only if $a \vee b$.
\begin{proof}
If $a \vee b$ then trivially $a \cup b$, $a \cup b$ being an upper bound of $a$ and $b$.

The other direction is obtained by simple induction on $a$.
\end{proof}
\end{lemma}

$\Sier$ has a countable join operator, but it is limited to increasing sequences. Thanks to the binary join we can remove this limit to define interpret properties $\exists n : \mathbb{N}, P_n$ and even $\exists x : A, P\; x$ when $A$ is enumerable.

\begin{definition} \label{def:sier-countable-join}
For all $f : \mathbb{N} \rightarrow \Sier$ there is a least upper bound $\sup f$ of all the $f_n$.
\begin{proof}
We have $\sup f$ for monotonous sequences, so for arbitrary $f : \mathbb{N} \rightarrow \Sier$ we define $f^\leq : \mathbb{N} \rightarrow \Sier$ by $f^\leq\; n \coloneqq \bigcup_{m \leq n} f\; m$.

Then $f^\leq$ is monotonous and $\sup f \coloneqq \sup f^\leq$ is the least upper bound of all the $f_n$.
\end{proof}
\end{definition}

That $\sup f$ semi-decides $\exists x, f\; x$ is trivial.

\subsection{Interleaving}

\begin{definition}[Disjoint] \label{def:disjoint}
$a$ and $b : \Sier$ are disjoint when they do not both hold, i.e. $a \rightarrow b \rightarrow \Empty$.
\end{definition}

Interleaving lets us define a value in $\Bool_\bot$ from two values in $\Sier$ which are not both $\top$. If we see $x\; y : \Sier$ as semi-decision procedures then the interleaving of $x$ and $y$ is $\Eta true$ if $x$ terminates (i.e. $x = \top$), $false$ if $y$ terminates and does not terminate if neither terminates. If computing on a Turing machine it would be obtained by interleaving simulated steps of $x$ and $y$ until one terminates, then returning a value depending on which one terminated.

We can only interleave disjoint values: a Turing machine could pick whichever one terminates first, but we have hidden those distinctions away using higher inductive types.

\begin{definition} \label{def:interleave}
We define by mutual induction a function
\begin{equation*}
\begin{split}
interleave\star : \forall a\; b : \Sier, disjoint\; a\; b \rightarrow \\
	\Sigma_{c : \Sier} (map\; (\lambda \_, false)\; b) \leq c)
\end{split}
\end{equation*}
where $map : \forall A \; B : Type, (A \rightarrow B) \rightarrow A_\bot \rightarrow B_\bot$ is the map of the partiality monad, and in parallel a proof that for all $a\; a' : \Sier$, if $a \leq a'$ then for all $b : \Sier$ disjoint with $a$ and with $a'$, $interleave_\star\; a\; b \leq interleave_\star\; a'\; b$.

Then the interleaving function $interleave$ is the first projection of $interleave_\star$. It computes as follows:
\begin{itemize}
\item $interleave\; \top\; b \coloneqq \Eta true$
\item $interleave\; \bot\; b \coloneqq map\; (\lambda \_, false) b$
\item $interleave\; (\sup f)\; b \coloneqq \sup (\lambda n, interleave\; f_n\; b$
\end{itemize}
Some attention must be taken to keep track of the disjointness proofs which are left implicit on paper.
\end{definition}

\begin{lemma} \label{lem:interleave-top-r}
If $a : \Sier$ is disjoint from $\top$ then
\[ interleave\; a\; \top = \Eta false \]
\begin{proof}
$a$ is disjoint from $\top$ so $a = \bot$ and $interleave\; a\; \top = map\; (\lambda \_, false)\; \top = \Eta false$.
\end{proof}
\end{lemma}

\begin{lemma} \label{lem:interleave-pr}
For $a\; b : \Sier$ disjoint $interleave\; a\; b = \Eta true$ (resp. $\Eta false$) if and only if $a$ holds (resp. $b$ holds).
\begin{proof}
By simple induction on $a$ in the first direction, by computation in the second (note that if $b$ then $a = \bot$ as they are disjoint).
\end{proof}
\end{lemma}

\subsection{Partial comparison of real numbers with rational numbers}

\begin{lemma} \label{lem:semidecidable-compare-rat}
For all $x : \real$ and $q : \mathbb{Q}$, $x < \Rat q$ is semi-decidable, i.e. $\exists s : \Sier, s \leftrightarrow x < \Rat q$.
\begin{proof}
By simple induction on $x$.

In the base case, for all $q\; r : \mathbb{Q}$, $\Rat q < \Rat r$ is decidable so we pick $s \coloneqq \top$ or $\bot$ as appropriate.

In the limit case, if $x : \Approximation \real$ such that for all $\varepsilon$ and $q$, $x_\varepsilon < \Rat q$ is semi-decidable, let $q : \mathbb{Q}$, we take $s \coloneqq \exists \varepsilon\; \delta : \mathbb{Q}^+, x_\varepsilon < \Rat (q - \varepsilon - \delta)$ (interpreted as a value in $\Sier$). Then to show correctness:
\begin{itemize}
\item if $\exists \varepsilon\; \delta : \mathbb{Q}^+, x_\varepsilon < \Rat (q - \varepsilon - \delta)$ then $\lim x < \Rat q = \Rat (q - \varepsilon - \delta + \varepsilon + \delta)$ by \cref{lem:r-lt-close-plus}.
\item if $\lim x < \Rat q$, there merely is $r : \mathbb{Q}$ such that $\lim x < \Rat r$ and $r < q$. Let $\varepsilon \coloneqq q - r$.\\
	Then $x_{\frac{\varepsilon}{4}} < \Rat (q - \varepsilon - \varepsilon) = \Rat (r + \varepsilon + \varepsilon)$ by \cref{lem:r-lt-close-plus}.
\end{itemize}
\end{proof}
\end{lemma}

\begin{definition} \label{def:is-positive}
For $x : \real$ let $isPositive\; x$ be the interleaving of the semi-decisions for $- x < 0$ and $x < 0$.
\end{definition}

\begin{theorem} \label{thm:is-positive-ok}
Let $x : \real$.
\begin{itemize}
\item $0 < x$ if and only if $isPositive\; x = \Eta true$
\item $x < 0$ if and only if $isPositive\; x = \Eta false$
\item $isPositive\; 0 = \bot$
\end{itemize}
\begin{proof}
By \cref{lem:interleave-top-r,lem:interleave-pr} and computation.
\end{proof}
\end{theorem}

\section{Conclusion} \label{sec:conclusion}

We have defined a Cauchy completion operation which is a monad on the category of spaces with an appropriate closeness relation and Lipschitz functions. When applied to the space of rational numbers it produces a Cauchy complete archimedean ordered field generated by rationals and limits of Cauchy approximations, i.e. the Cauchy reals. Finally we have defined and proven correct a semi-decision procedure (in the sense of \citet{Partiality}) for comparing a Cauchy real and a rational number.

\section*{Acknowledgements}
\addcontentsline{toc}{section}{Acknowledgements}

This paper is the result of two internships under the direction of Bas Spitters, at Radbough University and at the University of Aarhus.


\bibliographystyle{abbrvnat}


\end{document}